\documentclass[12pt]{iopart}
\usepackage{graphicx}
\usepackage{latexsym}
\usepackage{verbatim}

\begin{document}

\title{Bell's inequalities in the tomographic representation}

\author{C. Lupo$^1$, V. I. Man'ko$^2$, G. Marmo$^1$}
\address{$^1$ Dipartimento di Scienze Fisiche, Universit\'a "Federico II" e sezione INFN di Napoli,
Complesso Universitario di Monte Sant'Angelo, via Cintia, 80126
Napoli, Italy}
\address{$^2$ P. N. Lebedev Physical Institute, Leninskii Prospect 53, Moscow 119991, Russia}
\ead{\mailto{lupo@na.infn.it}, \mailto{manko@sci.lebedev.ru},
\mailto{marmo@na.infn.it}}

\begin{abstract}

The tomographic approach to quantum mechanics is revisited as a
direct tool to investigate violation of Bell-like inequalities.
Since quantum tomograms are well defined probability distributions,
the tomographic approach is emphasized to be the most natural one to
compare the predictions of classical and quantum theory. Examples of
inequalities for two qubits an two qutrits are considered in the
tomographic probability representation of spin states.

\end{abstract}



\section{Introduction}

Bell's inequalities were originally formulated \cite{Bell} in order
to provide a mathematical characterization of classical local hidden
variables theories. In their original formulation, Bell's
inequalities are propositions concerning expectation values of
dichotomic observables (such as spin$-1/2$ polarization), when two
spatially separated systems and local measurements are considered,
in presence of perfect (anti-) correlations between the two systems
relevant observables (such as two spin$-1/2$ in a singlet state).
The experimental violation of these inequalities is an evidence
against local classical variables models. Later on, other
inequalities were proposed that generalize the Bell's idea to the
case of non perfectly (anti-) correlated spin$-1/2$ systems
\cite{CHSH,CH}, to the case of spin of higher value \cite{Mermin}
and concerning probability of measurement output instead of
measurement expectation value \cite{Wigner}.

It is a remarkable fact that not all the states of a (say) bipartite
quantum system do violate some Bell-like inequalities: only states
that are \emph{entangled} are truly non local and not allowed to be
described by means of a \emph{classical} local variables model. With
the development of the theory of quantum information and in view of
the special role played by entangled states in quantum information
protocols, a violation of some Bell-like inequalities has assumed
also an operational role as a witness of entanglement. The power of
Bell-like inequalities is that they refer only to observables
quantities, as expectation value, correlations and probabilities
without an explicit link to the underlying theory. If a Bell-like
inequality is a proposition that is true for a classical theory, it
is nevertheless a well defined proposition (not necessary true) in
the framework of quantum theory. Hence the very idea of Bell's
inequalities leads to consider a unified description of both
classical and quantum mechanics based on fundamental quantities as
probability distributions.

The conventional description of pure quantum states is by means of
wave functions \cite{Sch} or state vector in Hilbert space
\cite{Dirac}. For mixed states, the density matrix
\cite{Landau27,vonN} is used to describe quantum states. The problem
of measuring the quantum states was considered as the problem of
finding the Wigner function \cite{Wigner32}, by means of which the
optical tomograms of the states
\cite{Bertrand-Bertrand87,Vogel-Risken89}, which are the probability
distribution densities of the homodyne photon quadratures, can be
determined. In \cite{mancini.manko.tombesi} the use of symplectic
tomogram as a tool for state reconstruction was extended in order to
describe the quantum state by the probability distribution from the
very beginning. This approach is called "tomographic probability
representation of quantum states". For spin degrees of freedom the
probability representation was found in
\cite{discrete.variables,OlgaJEPT} for one qudit and in
\cite{Manko_2} for two qudits. In the framework of the tomographic
representation, the spin state is identified with the probability
distribution of spin projection on direction labeled by angular
coordinates on the Bloch spheres for arbitrary number of qudits.

The tomographic map from state vectors or density matrices onto fair
probability distributions contains complete information on the
quantum states. Its mathematical structure was recently found in
\cite{ventriglia}. The relation of tomographic probability
representation with the star-product quantization procedure was
established in \cite{kernel}.

The aim of this work is to find new explicit formulas for spin
tomograms of two qubits and two qutrits and to analyze, by means of
these formulas, some Bell-like inequalities. The paper is organized
as follows. In section \ref{tomo.sep} we review the separability
problem using the tomographic probability description of spin
states. In section \ref{qubits} we derive the formulas for spin
tomograms of two qubits and study the CHSH inequalities \cite{CHSH}.
In section \ref{qutrits} we obtain the probability representation
for multiqutrit state. In section \ref{conclusions} we present the
conclusions.

\section{Tomograms and separability}\label{tomo.sep}

A tomographic description of quantum system can be formulated for
systems with both discrete and continuous variables
\cite{ventriglia}. Here we are interested in the case of discrete
variable systems that we are going to describe in the framework of
spin tomography.

For qudit states with spin $j$ the tomographic probability
distribution is defined as the diagonal elements of the density
operator
\begin{equation}
\rho_U = U^\dag \rho U
\end{equation}
in a standard basis $\{ |m\rangle \}_{m=-j,\dots j}$, where $U$ is
an operator of the unitary irreducible representation of the
$\mathrm{SU}(2)$ group. The tomogram of the qudit state reads
\begin{equation}
\omega(m,\stackrel{\rightarrow}{n}) = \langle m | \rho_U | m \rangle
= \langle m | U^\dag \rho U | m \rangle\;.
\end{equation}
Here $\stackrel{\rightarrow}{n} =
(\sin{\theta}\cos{\phi},\sin{\theta}\sin{\phi},\cos{\theta})$ is an
unit vector determining a point on the Bloch sphere. The tomogram
is, by construction, the probability distribution of the spin
projection $m$ onto the direction $\stackrel{\rightarrow}{n}$. The
probability distribution determines the density matrix $\rho$. The
formula connecting the tomogram
$\omega(m,\stackrel{\rightarrow}{n})$ with the density matrix $\rho$
was obtained in \cite{OlgaJEPT}. For example the tomographic
probability of the qubit state
\begin{eqnarray}
\rho = \left[
\begin{array}{cc}
1 & 0 \\
0 & 0
\end{array}\right]
\end{eqnarray}
reads as follows
\begin{eqnarray}
\omega(1/2,\stackrel{\rightarrow}{n}) & = & \cos^2{\theta/2}\;, \\
\omega(-1/2,\stackrel{\rightarrow}{n}) & = & \sin^2{\theta/2}\;.
\end{eqnarray}
We used the matrix $U$ rotating the spinor in the form
\begin{eqnarray}
U = \left[
\begin{array}{cc}
\cos{\theta/2}e^{i\frac{\phi+\psi}{2}} &
\sin{\theta/2}e^{i\frac{\phi-\psi}{2}} \\
-\sin{\theta/2}e^{-i\frac{\phi-\psi}{2}} &
\cos{\theta/2}e^{-i\frac{\phi+\psi}{2}}
\end{array}\right]\;.
\end{eqnarray}
Here $\phi,\theta,\psi$ are the Euler angles. For two qudits the
tomogram is defined as follows:
\begin{equation}
\omega(m_1,m_2;\stackrel{\rightarrow}{n_1},\stackrel{\rightarrow}{n_2})
= \langle m_1 m_2 | \mathcal{U}^\dag \rho \mathcal{U} | m_1 m_2
\rangle\;,
\end{equation}
where $\rho$ is a density matrix of two qudits, $\mathcal{U}=U_1
\otimes U_2$, and the matrices $U_1$ and $U_2$ are matrices of
irreducible representation of the group $\mathrm{SU}(2)$
corresponding to the first and second qudit, respectively. The spin
projections $m_1$ and $m_2$ onto directions
$\stackrel{\rightarrow}{n_1}$ and $\stackrel{\rightarrow}{n_2}$ are
random variables of the tomogram which is joint probability
distribution function for the two spin projections. Below we discuss
in more details the generic qudits tomograms.

Let us consider an operator $A^{(j)}$ acting on a space of a
spin$-j$ irreducible representation of $\mathrm{SU}(2)$. Given a
standard basis $\{ |jm\rangle \}$ with $m=-j,-j+1,...j-1,j$ the
matrix elements of the operator
\begin{equation}
A_{m,m'}^{(j)} = \langle m | A^{(j)} | m' \rangle
\end{equation}
of course completely determine the operator
\begin{equation}
A^{(j)} = \sum A^{(j)}_{m,m'} | m \rangle\langle m' |\;.
\end{equation}
We consider the diagonal elements in a rotated frame
\begin{equation} \label{tomog}
\omega_A(m,\Omega) = \langle m | R^\dag(\Omega) A^{(j)} R(\Omega) |
m \rangle = \tr\left[ A^{(j)} R(\Omega) |m \rangle\langle m|
R^\dag(\Omega) \right]\;,
\end{equation}
where $R(\Omega)$ is a unitary spin$-j$ representation of
$\mathrm{SU}(2)$ and $\Omega$ is a short hand notation for the three
Euler angles $\alpha$, $\beta$ and $\gamma$. The diagonal elements,
as functions of the variable $m$ and of the parameters $\Omega$
define the spin tomogram of the operator $A^{(j)}$. In the case in
which $A^{(j)}$ represents a density operator describing the state
of a spin$-j$ system, the tomogram $\omega_A(m,\Omega)$ is
interpreted as the probability of finding the system with
polarization $m$ along the $z$ axis in a system rotated with Euler
angles $\Omega$. The tomogram (\ref{tomog}) is a family of well
defined probability distribution on the variable $m$ with parameter
$\stackrel{\rightarrow}{n}$:
\begin{eqnarray}
\omega_A(m,\stackrel{\rightarrow}{n}) & \geq & 0\;,\\
\sum_m \omega_A(m,\stackrel{\rightarrow}{n}) & = & 1\;.
\end{eqnarray}

It is a remarkable result that the knowledge of only diagonal matrix
elements in a generic rotated frame is sufficient to reconstruct the
operator:
\begin{equation}
A^{(j)} = \sum_{m=-j}^{j} \int d\Omega K(m,\Omega)
\omega_A(m,\Omega)\;,
\end{equation}
where
\begin{equation}
\int d\Omega = \int_0^{2\pi} d\alpha \int_0^\pi \sin{\beta}d\beta
\int_0^{2\pi} d\gamma\;.
\end{equation}
The explicit expression for the \emph{quantizer} operator
$K(m,\Omega)$ was found in \cite{kernel}.

Notice that as long as the polarization along the $z$ axis is
considered, the spin tomogram (\ref{tomog}) depends only on two
Euler angles: in the following we write
\begin{equation}
\Pi^{(j)}(m,\stackrel{\rightarrow}{n}) = R(\Omega) |m \rangle\langle
m| R^\dag(\Omega)\;,
\end{equation}
where $\stackrel{\rightarrow}{n} =
(\cos{\alpha}\sin{\beta},\sin{\alpha}\sin{\beta},\cos{\beta})$ is
the rotated axis of polarization. Hence, in the tomographic
approach, the state of a quantum system is described by means of a
well defined probability distribution
$\omega(m,\stackrel{\rightarrow}{n})$ related to a Stern
Gerlach-like measurement along the direction
$\stackrel{\rightarrow}{n}$. Notice that a Bloch sphere description
is obtained for the quantum state even for $j > 1/2$.

One of the open problems in quantum mechanics and quantum
information theory is to give a complete characterization of
entangled states. Given a bipartite system, a quantum state of the
system is said to be separable if it can be written as a convex sum
of factorized states:
\begin{equation} \label{def-sep}
\rho = \sum_k p_k \rho_k^{(A)} \otimes \rho_k^{(B)}\;, \ \ \sum_k
p_k = 1\;.
\end{equation}
Otherwise the state is said to be entangled. Let us also recall that
a factorized state $\rho = \rho^{(A)}\otimes\rho^{(B)}$ is called a
simply separable state. These definitions can be generalized, with
some care, to the case of multi-partite systems \cite{Cirac,Lupo}.

The relation between local realism and separability of quantum
states was widely studied. It is clear from the definition
(\ref{def-sep}), that every separable states can be described by
means of a local hidden variables model (where the hidden variable
can be identified with the index $k$). In \cite{Werner} it was first
shown with an example that the converse is not true, \emph{i.e.}\
there exist quantum states that can be described by a hidden local
variables model but are nevertheless entangled. This means that the
violation of a Bell's inequalities by a given quantum states is a
sufficient (though not necessary) condition for the state to be
entangled. Although a systematic approach to generate all Bell's
inequalities exists \cite{Pitowski}, how to find the inequality that
presents a maximal violation for a given entangled state is still an
open problem.

From the point of view of entanglement detection and
characterization, it is interesting to consider the tomographic
description of state of multipartite quantum systems. To fix the
ideas, let us consider a bipartite system composed of one spin$-j_1$
and one spin$-j_2$: in this case the spin tomogram of a state of the
compound system described by density matrix $\rho$ is written as
follows:
\begin{equation}
\omega_\rho(m_1,m_2;\stackrel{\rightarrow}{n_1},\stackrel{\rightarrow}{n_2})
= tr\left( \rho
\Pi^{(j_1)}(m_1,\stackrel{\rightarrow}{n_1})\otimes\Pi^{(j_2)}(m_2,\stackrel{\rightarrow}{n_2})
\right)\;.
\end{equation}
This definition is simply generalized to the case of multipartite
spin systems and refers to local Stern Gerlach-like measurement.

For example the tomographic probability distribution function for
the two qubit state
\begin{eqnarray}
\rho = \left[
\begin{array}{cccc}
1 & 0 & 0 & 0 \\
0 & 0 & 0 & 0 \\
0 & 0 & 0 & 0 \\
0 & 0 & 0 & 0
\end{array}\right]
\end{eqnarray}
reads
\begin{eqnarray}
\omega(1/2,1/2;\stackrel{\rightarrow}{n_1},\stackrel{\rightarrow}{n_2})   & = & \cos^2{\theta_1/2} \cos^2{\theta_2/2}\;, \\
\omega(1/2,-1/2;\stackrel{\rightarrow}{n_1},\stackrel{\rightarrow}{n_2})  & = & \cos^2{\theta_1/2} \sin^2{\theta_2/2}\;, \\
\omega(-1/2,1/2;\stackrel{\rightarrow}{n_1},\stackrel{\rightarrow}{n_2})  & = & \sin^2{\theta_1/2} \cos^2{\theta_2/2}\;, \\
\omega(-1/2,-1/2;\stackrel{\rightarrow}{n_1},\stackrel{\rightarrow}{n_2})
& = & \sin^2{\theta_1/2} \sin^2{\theta_2/2}\;.
\end{eqnarray}
The state is simply separable and the tomographic probability has
the form of factorized joint probability distribution
\begin{equation}
\omega(m_1,m_2;\stackrel{\rightarrow}{n_1},\stackrel{\rightarrow}{n_2})
= \omega_1(m_1,\stackrel{\rightarrow}{n_1})
\omega_2(m_2,\stackrel{\rightarrow}{n_2})\;,
\end{equation}
where the probability distributions $\omega_1$ and $\omega_2$
describe the states of the first and second spin respectively. The
joint tomographic probability determines the density matrix by means
of inversion formula obtained in \cite{Manko_2}. Due to linearity of
the tomographic map of density matrices onto joint probability
distributions of spin projections, the tomogram of a separable state
is the convex sum of factorized joint probability distributions of
the simply separable states:
\begin{equation}
\omega(m_1,m_2;\stackrel{\rightarrow}{n_1},\stackrel{\rightarrow}{n_2})
= \sum_k p_k \omega_1^{(k)}(m_1,\stackrel{\rightarrow}{n_1})
\omega_2^{(k)}(m_2,\stackrel{\rightarrow}{n_2})\;.
\end{equation}

\section{Qubits tomograms}\label{qubits}

In this section we discuss the tomographic representation for
spin$-1/2$ (qubit) systems in its link with standard density matrix
description. Let us first consider a one-qubit system. It is well
known that a qubit density state can be written in terms of Pauli
matrices:
\begin{equation}
\rho_1 = \frac{1}{2} \left( \sigma_0 + x^i \sigma_i \right)\;,
\end{equation}
where (the sum over repeated indices is intended)
\begin{equation} \label{trace}
x^i = \delta^{ij} \tr(\rho \sigma_j) = \delta^{ij} x_j
\end{equation}
since
\begin{equation}
\tr(\sigma_i \sigma_j) = 2\delta_{ij}\;.
\end{equation}

In the following we take $m=-1,1$. With this convention, from the
definition (\ref{tomog}) it follows that in the tomographic
representation:
\begin{equation} \label{1-qubit-tomo}
\omega(m,\stackrel{\rightarrow}{n}) = \tr\left( \rho_1
\Pi(m,\stackrel{\rightarrow}{n}) \right)\;,
\end{equation}
where
\begin{equation}
\Pi(m,\stackrel{\rightarrow}{n}) = \frac{1}{2} \left( \sigma_0 + m
n^i \sigma_i \right)
\end{equation}
is the projector on the eigenstate with polarization $m$ along the
direction $\stackrel{\rightarrow}{n}=(n^1,n^2,n^3)$, where for
convenience we have chosen $m=\pm 1$. The operator
$\Pi(m,\stackrel{\rightarrow}{n})$ plays the role of the
de-quantizer operator used in star-product quantization scheme
\cite{star-prod}.

From (\ref{1-qubit-tomo}) and (\ref{trace}) it follows that the
explicit expression for a generic qubit tomogram is
\begin{equation}
\omega_1(m,\stackrel{\rightarrow}{n}) = \frac{1}{2} \left( 1 + m
\stackrel{\rightarrow}{n} \cdot \stackrel{\rightarrow}{x} \right)\;,
\end{equation}
where $\stackrel{\rightarrow}{x}=(x_1,x_2,x_3)$ and
$\stackrel{\rightarrow}{n} \cdot \stackrel{\rightarrow}{x} = n^i
x_i$. The expression (\ref{1-qubit-tomo}) can be immediately
generalized to the case of multi-qubit system. In the case of a
system of $N$ qubits in a global state $\rho_N$, the (global)
tomogram is given by the following relation:
\begin{equation} \label{N-qubit-tomo}
\omega_N(m_1,m_2,\dots m_N;\stackrel{\rightarrow}{n_1},
\stackrel{\rightarrow}{n_2},\dots \stackrel{\rightarrow}{n_N}) =
\tr\left[ \rho_N \bigotimes_{i=1...N}
\Pi(m_i,\stackrel{\rightarrow}{n_i})\right]\;.
\end{equation}
In the case of a system of two qubits (\ref{N-qubit-tomo})
simplifies to
\begin{equation} \label{2-qubit-tomo}
\omega_2(m_1,m_2;\stackrel{\rightarrow}{n_1},\stackrel{\rightarrow}{n_2})
= \tr\left[ \rho_2 \frac{1}{4}(\sigma_0 + m_1 n_1^i \sigma_i)
\otimes (\tau_0 + m_2 n_2^i \tau_i) \right]\;,
\end{equation}
where $\sigma_\mu$ and $\tau_\mu$ are the Pauli matrices
respectively related to the first and second qubit.

Defining $x_i = \tr(\rho_2 \sigma_i)$, $y_i = \tr(\rho_2 \tau_i)$
and $z_{ij} = \tr(\rho_2 \sigma_i\otimes\tau_j)$, where $\sigma_i$
and $\tau_i$ are short-hand notation for $\sigma_i\otimes\tau_0$ and
$\sigma_0\otimes\tau_i$ respectively, the tomogram
(\ref{2-qubit-tomo}) reads:
\begin{equation}\label{2-qubit-tomo-function}
\omega(m_1,m_2;\stackrel{\rightarrow}{n_1},\stackrel{\rightarrow}{n_2})
= \frac{1}{4} \left( 1 + m_1 n_1^i x_i + m_2 n_2^i y_i + m_1 m_2
n_1^i z_{ij} n_2^j \right)\;.
\end{equation}
Notice that for simply separable states $\tr(\rho_2 \sigma_i \otimes
\tau_j) = \tr(\rho_2 \sigma_i) \tr(\rho_2 \tau_j)$, \emph{i.e.}\
$z_{ij} = x_i y_j$ and the tomogram assumes a factorized form:
\begin{equation}
\omega(m_1,m_2;\stackrel{\rightarrow}{n_1},
\stackrel{\rightarrow}{n_2}) = \frac{1}{4} \left( 1 + m_1
\stackrel{\rightarrow}{n_1}\cdot\stackrel{\rightarrow}{x} \right)
\left( 1 + m_2
\stackrel{\rightarrow}{n_2}\cdot\stackrel{\rightarrow}{y} \right)\;.
\end{equation}

\subsection{Two spin$-1/2$ Bell-Wigner inequalities}

Let us consider the inequality proposed in \cite{Wigner}. It is
related to the case of two spin$-1/2$ particles with perfect
anti-correlation. For each particle the polarization is
independently measured along three arbitrary directions. The joint
probability of finding the first and the second particles polarized
respectively in the $\stackrel{\rightarrow}{n_1}$ and
$\stackrel{\rightarrow}{n_2}$ direction is indicated with
$P(\stackrel{\rightarrow}{n_1},\stackrel{\rightarrow}{n_2})$. The
hypothesis of perfect anti correlation implies that the probability
of measure parallel polarization along a fixed direction vanishes:
\begin{equation}
P(\stackrel{\rightarrow}{n},\stackrel{\rightarrow}{n}) = 0\;.
\end{equation}
Given three arbitrary directions $\stackrel{\rightarrow}{n_a}$,
$\stackrel{\rightarrow}{n_b}$ and $\stackrel{\rightarrow}{n_c}$ the
following inequality holds for a classically correlated state
\cite{Wigner}:
\begin{equation} \label{W-ineq}
P(\stackrel{\rightarrow}{n_a},\stackrel{\rightarrow}{n_b}) +
P(\stackrel{\rightarrow}{n_b},\stackrel{\rightarrow}{n_c}) -
P(\stackrel{\rightarrow}{n_a},\stackrel{\rightarrow}{n_c}) \geq 0\;.
\end{equation}
Notice that these probability distributions are directly given in
the tomographic representation, since
\begin{equation}
P(\stackrel{\rightarrow}{n_1},\stackrel{\rightarrow}{n_2}) =
\omega(1,1; \stackrel{\rightarrow}{n_1},
\stackrel{\rightarrow}{n_2})\;.
\end{equation}

Inequality (\ref{W-ineq}) is obtained for perfectly classically
anti-correlated states. It is easy to see that a quantum simply
separable state cannot exhibit perfect (anti-) correlations, hence
we consider non-perfect anti-correlation in a simply separable state
of the following form:
\begin{equation}
\omega(m_1,m_2;\stackrel{\rightarrow}{n_1},
\stackrel{\rightarrow}{n_2}) = \frac{1}{4} \left[ 1 + m_1
(\stackrel{\rightarrow}{n_1}\cdot\stackrel{\rightarrow}{x}) \right]
\left[1 - m_2
(\stackrel{\rightarrow}{n_2}\cdot\stackrel{\rightarrow}{x})
\right]\;.
\end{equation}
For such a state (\ref{W-ineq}) are always fulfilled and are simply
written as follows:
\begin{eqnarray}
\omega(1,1;\stackrel{\rightarrow}{n_a},\stackrel{\rightarrow}{n_b})
+
\omega(1,1;\stackrel{\rightarrow}{n_b},\stackrel{\rightarrow}{n_c})
-
\omega(1,1;\stackrel{\rightarrow}{n_a},\stackrel{\rightarrow}{n_c}) =  \\
\frac{1}{4}\left[ 1 -
(\stackrel{\rightarrow}{n_a}\cdot\stackrel{\rightarrow}{x})
(\stackrel{\rightarrow}{n_b}\cdot\stackrel{\rightarrow}{x}) -
(\stackrel{\rightarrow}{n_b}\cdot\stackrel{\rightarrow}{x})
(\stackrel{\rightarrow}{n_c}\cdot\stackrel{\rightarrow}{x}) +
(\stackrel{\rightarrow}{n_a}\cdot\stackrel{\rightarrow}{x})
(\stackrel{\rightarrow}{n_c}\cdot\stackrel{\rightarrow}{x}) \right]
\geq 0\;,
\end{eqnarray}
that is
\begin{equation}
(\stackrel{\rightarrow}{n_a}\cdot\stackrel{\rightarrow}{x})
(\stackrel{\rightarrow}{n_b}\cdot\stackrel{\rightarrow}{x}) +
(\stackrel{\rightarrow}{n_b}\cdot\stackrel{\rightarrow}{x})
(\stackrel{\rightarrow}{n_c}\cdot\stackrel{\rightarrow}{x}) -
(\stackrel{\rightarrow}{n_a}\cdot\stackrel{\rightarrow}{x})
(\stackrel{\rightarrow}{n_c}\cdot\stackrel{\rightarrow}{x}) \leq
1\;.
\end{equation}
Since the inequalities are fulfilled by non-perfectly
anti-correlated particles in a factorized state it follows that the
same is true for a generic anti-correlated separable state.

As a simple example, we consider the case of a two-qudit system in
the Werner state, defined for $\phi \in [-1,1]$ as follows:
\begin{equation} \label{Wer}
\rho_d(\phi) =  \frac{1}{d^3-d^2}\left[ (d-\phi) Id_{d^2} +
(d\phi-1) V \right]\;,
\end{equation}
where $Id_{d^2}$ is the identity operator in the compound system
space and $V$ is the swap operator ($V \psi\otimes\phi =
\phi\otimes\psi$). These states are symmetric under local unitary
operations of the kind $U \otimes U$: hence we expect a particular
simple tomographic expression for these states. The state
(\ref{Wer}) is known to be entangled for $\phi < 0$ and separable
otherwise. Notice that a spin$-j$ system can be viewed as a qudit
with $d=2j+1$.

In the case of two qubits ($d=2$) the tomogram of (\ref{Wer}) reads
as follows:
\begin{equation} \label{W-tomo}
\omega_W = \frac{1}{4} \left[ 1 + \frac{2\phi-1}{3} m_1 m_2 (
\stackrel{\rightarrow}{n_1} \cdot \stackrel{\rightarrow}{n_2} )
\right]\;.
\end{equation}
In terms of tomogram, the inequality (\ref{W-ineq}) is immediately
written as
\begin{equation}
\frac{2\phi-1}{3} \left[ (\stackrel{\rightarrow}{n_a} \cdot
\stackrel{\rightarrow}{n_c}) - (\stackrel{\rightarrow}{n_a} \cdot
\stackrel{\rightarrow}{n_b}) - (\stackrel{\rightarrow}{n_b} \cdot
\stackrel{\rightarrow}{n_c}) \right] \leq 1\;.
\end{equation}
It follows that the inequality (\ref{W-ineq}) is violated for any
$\phi < -1/2$.

\subsection{Two spin$-1/2$ CHSH inequalities}

As we have recalled above, both the Bell's inequalities \cite{Bell}
and Bell-Wigner inequalities \cite{Wigner} assume perfect (anti-)
correlations between the two system qubits. The inequalities known
as CHSH inequalities were introduced \cite{CHSH} in order to relax
the hypothesis of perfect correlation between the two systems. Also
in this case we deal with dichotomic observables. In the following
we consider the case of a composite system of two spin$-1/2$ and the
relevant observables are local magnetizations along a couple of
directions. As in the original Bell argument, but in contrast with
the Wigner approach, these inequalities are expressed in terms of
expectation values and correlations of local observables. Some
aspects of CHSH inequalities and their relation to tomographic
probabilities were discussed in \cite{Manko}.

Given two arbitrary directions $\stackrel{\rightarrow}{n_1}$ and
$\stackrel{\rightarrow}{n_2}$, let us consider the function
\begin{equation} \label{CHSH-corr}
M(\stackrel{\rightarrow}{n_1},\stackrel{\rightarrow}{n_2}) =
\tr(\rho_2 n_1^i\sigma_i \otimes n_2^j\tau_j)\;,
\end{equation}
that represents the correlation between the polarizations along the
$\stackrel{\rightarrow}{n_1}$ and $\stackrel{\rightarrow}{n_2}$
direction, respectively for the first and second qubit, over the two
qubits density state $\rho_2$. Notice that, in terms of tomograms,
the correlation function (\ref{CHSH-corr}) can be easily written as
\begin{equation}
M(\stackrel{\rightarrow}{n_1},\stackrel{\rightarrow}{n_2}) =
\sum_{m_1,m_2} m_1 m_2
\omega(m_1,m_2;\stackrel{\rightarrow}{n_1},\stackrel{\rightarrow}{n_2})\;.
\end{equation}

Given four arbitrary directions $\stackrel{\rightarrow}{n_a}$,
$\stackrel{\rightarrow}{n_b}$, $\stackrel{\rightarrow}{n_c}$ and
$\stackrel{\rightarrow}{n_{b'}}$, the CHSH inequalities read as
follows:
\begin{equation} \label{CHSH-ineq}
| M(\stackrel{\rightarrow}{n_a},\stackrel{\rightarrow}{n_b}) -
M(\stackrel{\rightarrow}{n_a},\stackrel{\rightarrow}{n_c}) | +
M(\stackrel{\rightarrow}{n_{b'}},\stackrel{\rightarrow}{n_b}) +
M(\stackrel{\rightarrow}{n_{b'}},\stackrel{\rightarrow}{n_c}) - 2
\leq 0\;.
\end{equation}

For two qubits Werner state, using (\ref{W-tomo}), the average
magnetization is easily written as
\begin{equation}
M(\stackrel{\rightarrow}{n_1},\stackrel{\rightarrow}{n_2}) =
\frac{2\phi-1}{3} \left( \stackrel{\rightarrow}{n_1} \cdot
\stackrel{\rightarrow}{n_2} \right)\;.
\end{equation}
The inequality (\ref{CHSH-ineq}) reads
\begin{equation}
\frac{|2\phi-1|}{3} \left[ |
\stackrel{\rightarrow}{n_a}\cdot(\stackrel{\rightarrow}{n_b} -
\stackrel{\rightarrow}{n_c}) | -
\stackrel{\rightarrow}{n_{b'}}\cdot(\stackrel{\rightarrow}{n_b} +
\stackrel{\rightarrow}{n_c}) \right] \leq 2\;.
\end{equation}
Notice that the maximum of the function
\begin{equation}
Y(\stackrel{\rightarrow}{n_a},\stackrel{\rightarrow}{n_b},\stackrel{\rightarrow}{n_{b'}},\stackrel{\rightarrow}{n_c})
=  | \stackrel{\rightarrow}{n_a}\cdot (\stackrel{\rightarrow}{n_b} -
\stackrel{\rightarrow}{n_c}) | - \stackrel{\rightarrow}{n_{b'}}\cdot
(\stackrel{\rightarrow}{n_b} + \stackrel{\rightarrow}{n_c})
\end{equation}
is reached when
\begin{eqnarray}
\stackrel{\rightarrow}{n_a}    &=& \pm\frac{\stackrel{\rightarrow}{n_b}-\stackrel{\rightarrow}{n_c}}{|\stackrel{\rightarrow}{n_b}-\stackrel{\rightarrow}{n_c}|}  \\
\stackrel{\rightarrow}{n_{b'}} &=& -
\frac{\stackrel{\rightarrow}{n_b}+\stackrel{\rightarrow}{n_c}}{|\stackrel{\rightarrow}{n_b}+\stackrel{\rightarrow}{n_c}|}
\end{eqnarray}
and $\stackrel{\rightarrow}{n_b} \cdot \stackrel{\rightarrow}{n_c} =
0$, and it is equal to $2\sqrt{2}$. The inequality is violated for
any $\phi < -\frac{3\sqrt{2}-2}{4}$. Hence, the violation of the
inequality does not detect entanglement when $-1 \leq \phi \leq
-\frac{3\sqrt{2}-2}{4}$.

\section{Qutrits tomography}\label{qutrits}

In the previous sections we were dealing with qubit systems. Let us
now consider the case of qutrits. In order to write the spin
tomogram for a generic qutrit state, one has to consider the $s=1$
irreducible representations of the group $\mathrm{SU}(2)$. Let us
consider a realization of the angular momentum as qutrits operators
$J_1, J_2, J_3$, such that
\begin{equation}
[ J_i , J_j ] = i \epsilon_{ij}^k J_k\;.
\end{equation}
In terms of this given representation, the spin tomogram of a qutrit
state is related to the standard density matrix description via the
following relation:
\begin{equation} \label{1-qutrit-tomo}
\omega(m,\stackrel{\rightarrow}{n}) = \tr( \rho_1
\Pi(m,\stackrel{\rightarrow}{n}) )\;,
\end{equation}
where $m=-1,0,1$, and the qutrit \emph{de-quantizer} operator is now
given by
\begin{equation}
\Pi(m,\stackrel{\rightarrow}{n}) = \left(1-m^2\right)Id_3 +
\frac{m}{2} n^i J_i + \left(\frac{3}{2} m^2 -1\right)(n^i J_i)^2\;,
\end{equation}
where $Id_3$ is the qutrit identity operator, and
$\Pi(m,\stackrel{\rightarrow}{n})$ is the projector on the
eigenvector of polarization $m$ along the
$\stackrel{\rightarrow}{n}$ direction. The relation
(\ref{1-qutrit-tomo}) is easily generalized in the case of a system
of $N$ qutrits as follows:
\begin{equation}
\omega(m_1,m_2,\dots
m_N;\stackrel{\rightarrow}{n_1},\stackrel{\rightarrow}{n_2},\dots
\stackrel{\rightarrow}{n_N}) = \tr\left[ \rho_N \bigotimes_{i=1...N}
\Pi(m_i,\stackrel{\rightarrow}{n_i})\right]\;.
\end{equation}

As an example let us consider the two-qutrits Werner state obtained
from (\ref{Wer}) with $d=3$:
\begin{equation} \label{Werner-state}
\rho_W = \frac{3-\phi}{24} Id_9 + \frac{3\phi-1}{24} V\;.
\end{equation}
The tomographic representation is explicitly given by
\begin{equation}
\omega(m_1,m_2;\stackrel{\rightarrow}{n_1},\stackrel{\rightarrow}{n_2})
= \tr\left[ \rho_W
\Pi(m_1,\stackrel{\rightarrow}{n_1})\otimes\Pi(m_2,\stackrel{\rightarrow}{n_2})
\right]\;,
\end{equation}
that yields to
\begin{eqnarray} \label{2-qutrit-tomo}
\omega(m_1, m_2 ; \stackrel{\rightarrow}{n_1} ,
\stackrel{\rightarrow}{n_2}) &=& \frac{3-\phi}{24} +
                                      \frac{3\phi-1}{24}\left[ 3\left(1-m_1^2\right)\left(1-m_2^2\right) \right.            \nonumber \\
                                 &+& \left(1-m_1^2\right)\left(3m_2^2-2\right) + \left(1-m_2^2\right)\left(3m_1^2-2\right)  \nonumber \\
                                 &+&  \frac{m_1 m_2}{2} (\stackrel{\rightarrow}{n_1} \cdot
                                      \stackrel{\rightarrow}{n_2})                                                          \nonumber \\
                                 &+& \left. \left(\frac{3}{2} m_1^2 -1\right)\left(\frac{3}{2} m_2^2
                                      -1\right)\left(1 + (\stackrel{\rightarrow}{n_1} \cdot \stackrel{\rightarrow}{n_2})^2\right)
                                      \right]\;.
\end{eqnarray}

As another example, we discuss the non-linear Bell-like inequality
proposed in \cite{Uffink}:
\begin{equation}
\langle A B' + A' B \rangle^2 + \langle A B - A' B' \rangle^2 \leq
1\;,
\end{equation}
where $A$, $A'$ and $B$, $B'$ are local observables for a system
composed of two spins, with the property of orthogonality $\tr(A
A')=0$, $\tr(B B')=0$. Although this inequality has been formulated
for a system of two qubits, it can be considered for a system of two
qutrits as well. If $A = n_A^i J_i$, $A' = n_{A'}^i J_i$, $B = n_B^i
J_i$, $B' = n_{B'}^i J_i$, from (\ref{2-qutrit-tomo}) we obtain that
\begin{equation}
\langle A B' \rangle = \frac{3\phi-1}{12}
\stackrel{\rightarrow}{n_A}\cdot\stackrel{\rightarrow}{n_{B'}}\;,
\end{equation}
and the inequality reads as follows:
\begin{equation}
\left[ \frac{3\phi-1}{24}\right]^2 \left[ \left(
\stackrel{\rightarrow}{n_A}\cdot\stackrel{\rightarrow}{n_{B'}} +
\stackrel{\rightarrow}{n_{A'}}\cdot\stackrel{\rightarrow}{n_B}
\right)^2 + \left(
\stackrel{\rightarrow}{n_A}\cdot\stackrel{\rightarrow}{n_B} -
\stackrel{\rightarrow}{n_{A'}}\cdot\stackrel{\rightarrow}{n_{B'}}\right)^2
\right] \leq 1\;.
\end{equation}
Notice that $\left[ \left(
\stackrel{\rightarrow}{n_A}\cdot\stackrel{\rightarrow}{n_{B'}} +
\stackrel{\rightarrow}{n_{A'}}\cdot\stackrel{\rightarrow}{n_B}
\right)^2 + \left(
\stackrel{\rightarrow}{n_A}\cdot\stackrel{\rightarrow}{n_B} -
\stackrel{\rightarrow}{n_{A'}}\cdot\stackrel{\rightarrow}{n_{B'}}\right)^2
\right]< 8$, therefore the inequality is never violated.

\section{Conclusions}\label{conclusions}

To conclude we point out the main results of the paper. We have
developed a formulation of Bell's inequalities by means of
tomographic probability distribution of spin projections describing
the quantum states completely. New formulas convenient for further
analysis for tomogram of one qubit, two qubits and tomograms of two
qutrits Werner state were obtained. The dequantizer operator for
qutrit is also a new result presented in the paper. We demonstrated
that both Wigner inequalities and CHSH inequalities as well their
violations can be easily explained using joint probability
distribution (tomograms) for spin projections. There are bounds for
the violation of CHSH inequalities discussed in
\cite{bounds,Tsirel,Popescu,bManko}. The CHSH inequalities
(\ref{CHSH-ineq}) are expressed exactly in terms of the function
(\ref{2-qubit-tomo-function}), it follows that the bound can be
found as the maximum of the left hand side of (\ref{CHSH-ineq}). We
will develop the analysis of Bell's inequalities based on
tomographic star-product approach in future publications.

\end{document}